\newcommand{\xo}{x_1}
\newcommand{\xt}{x_2}

\newcommand{\pbar}{\gamma^i p_i}
\newcommand{\ppbar}{\gamma^j p'_j}
\newcommand{\tvr}{\mid\mspace{-5mu}0'\mspace{-7mu}>}
\newcommand{\bvr}{\mid\mspace{-5mu}0 \mspace{-5mu}>}
\newcommand{\tvl}{<\mspace{-5mu}0'\mspace{-7mu}\mid}
\newcommand{\bvl}{<\mspace{-5mu}0 \mspace{-5mu} \mid}

\newcommand{\scbv}{<\mspace{-5mu}0\mspace{-5mu}  \mid \mspace{-5mu}0\mspace{-5mu}>}

\documentclass[11pt,a4paper]{article}
\usepackage[latin1]{inputenc}
\usepackage{amsmath}
\usepackage{amsfonts}
\usepackage{amssymb}
\begin{document}
\small

\title{The Contribution of Elelectron-Positron Pair Production
to the Vacuum Energy}

\author{Bernard R. Durney}
\date{2377 Route de Carc\'{e}s, F-83510 Lorgues, France durney@physics.arizona.edu} 

\maketitle


\smallskip\textbf{Abstract}.

The vacuum, defined as the state where no particles can be observed, is interpreted here to 
imply that the lifetime of the e-p pair should be equal to the Planck time, $t_P.$ Concerning
the title's subject, a perfect theory would require that the true vacuum expectation value of 
the operator associated with pair production, $S,$ be compatible with the normalization of
the true vacuum, $\tvr.$ At present, a calculation of the vacuum energy based on Feynman
diagrams reveals a serious difficulty: if only second order terms of the $S$-matrix are 
retained, and because there are no external lines, it follows that the space-time integrations 
over the coordinates $x_2,$ $x_1,$ required to calculate $\bvl S \bvr $, give rise to two 
identical delta functions. Therefore the $amplitude$ is proportional to the integration 
volume, $L^4,$ and, as a result, the square of the amplitude defies any physically meaningful 
interpretation. One is faced here with two evils:  modify the interaction Lagrangian 
so that the amplitude becomes proportional to $L^2;$ abstain from any calculation. 
It is felt that the first one is the lesser evil. If the $square$ of the amplitude is 
proportional to $L^4,$ this one can be interpreted as being the number, ${N}$, of events (pairs created), in the volume of integration. In the calculations for ${ N}$ it was assumed that the  
integral over momentums (rescaled to be dimensionless) was of the order unity, and that 
processes with small virtual photon-energy are predominant. The pairs' contribution to the 
vacuum density is then given by $\rho_V = 2 m{ N} t_P /VT, $ i.e., by the mass of the 
particles multiplied both, by the number of events per unit volume and time, and by the pairs' 
life time. For the value of $\rho_V$ it is found that $\rho_V  \approx 10^{-31}{\rm 
g\mspace{5mu} cm}^{-3}, $ in surprisingly good agreement with observations (in this type of 
calculations, powers of ten are epsilons).

\smallskip
\smallskip \textbf{1. Introduction.}
\smallskip
As a consequence of the commutation relations of creation and annihilation operators, $CR,$
the vacuum expectation value, $VEV,$ of the Hamiltonian of non interacting bosons- or fermions-
fields doesn't vanish, being positive and negative, respectively (c.f. for example Dolgov
et al. 1990, p. 112). It is usually argued that this outcome is related with 
Heisenberg's uncertainty principle. But this principle wouldn't demand a negative
$VEV$ for fermions. In fact, this result should rather be  an incentive to remain
open minded about the interpretation of this type of zero point energy. If the central issue
is the fulfilment of the uncertainty principle, it is conceivable that the non zero value of the $VEV$ is simply an indication that $physical$ $processes$ (to be distinguished from merely $CR$), leading to a non zero vacuum energy (as for example, particle pair production), are present in the concerned theory.
Incomparably more serious are the predictions of the vacuum energy generated by
symmetry breaking, c.f. Zee, 2010, p. 449. As Zee writes, "particle physics is built on a 
series of spontaneous symmetry breaking" each one contributing to the vacuum energy and
resulting in a gigantic discrepancy between theoretical expectations and observational reality.
Conceivably a solution to this problem will have to wait until the development of an 
"ultimate" quantum theory playing a	role for the microscopic world comparable
to the one played  by General Relativity for the macroscopic world.     
 
\smallskip
\smallskip \textbf{2. Calculation of the S matrix.}
\smallskip

In this paper,  Mandl and Shaw's  notation will be used, and citations, as number
of equations, figures...., will refer to their text, unless explicitly stated. The Feynman
diagram for pair creation, at $x_2,$ and annihilation, at $x_1$ is displayed in Fig 7.10, 
p.110 of their text book. In the calculations, c.g.s units will be used, and in the Appendix,
the values of some physical constants in these units are displayed. To second order the S 
matrix relevant to pair creation and annihilation is given by,
\begin{equation}
S = 2\pi i\alpha\int d^4 \xo d^4 x_2 S_F(x_2 - x_1) \gamma ^\alpha S_F(x_1 -x_2)
\gamma^\beta D^F_{\alpha \beta}(x_1 - x_2)
\end{equation}
where $\alpha$ is the fine structure constant and $S_F(x), D^F_{\alpha \beta}(x) $
are the Feynman's, fermion and photon propagators respectively. Their integral representation
is (c.f. Eqs (4.63), (5.27)),
\begin{equation}
S_F(x) = \frac{\hbar}{(2\pi\hbar)^4}\int d^4 p\mspace{5mu} e^{-ipx/\hbar} \frac{\pbar +mc}
{p^2 - (mc)^2 + i \epsilon}
\end{equation}
\begin{equation}
 D^F_{\alpha \beta}(x) = -\frac{g_{\alpha\beta}} {(2\pi)^4}\int d^4 k \mspace{5mu}
 \frac{e^{- ikx}}{k^2 + i \epsilon}
\end{equation}
With the help of these equations one obtains,
\begin{align}
\bvl S\bvr  = &-2\pi i \alpha \frac{\hbar^2}{(2\pi \hbar)^8}\frac{1} {(2\pi)^4}\int
d^4 x_1 d^4 x_2 d^4 p\mspace{5mu} d^4 p'\mspace{5mu} d^4 k \frac {\pbar + mc}{p^2 + (mc)^2 + 
i\epsilon}\gamma^\alpha \times\nonumber\\
& \times \frac {\ppbar + mc}{p'^2 + (mc)^2 + i\epsilon}\gamma^\beta g_{\alpha\beta}
\frac{1}{k^2+ i\epsilon}e^{i\xo(p/\hbar - p'/\hbar - k)}\mspace{5mu}
 e^{i\xt(p'/\hbar - p/\hbar + k)}
\end{align}
The integration over the coordinates $\xo, \xt,$ gives rise to two delta functions that are
identical, resulting in the  amplitude being proportional to $L^4$ which precludes 
any physically meaningful interpretation of the square of the amplitude (proportional
to $L^8).$ Here, $L^4 = c V T,$ where $V$ and $T$  are respectively, the volume and time of integration. 
As stated in the Abstract a perfect theory would require that the energy input
by the pairs be consistent with the true vacuum normalization. In the framework of Feynman
diagrams, and confined to work with the bare vacuum normalized to, $\scbv = 1,$ we would
like to interpret $\mid \bvl S\bvr \mid^2$ as the number of events (pair created)
in the volume of integration, $L^4.$ The amplitude must then be proportional
to $L^2.$ To achieve this goal, we have no other option left than to modify the 
interaction Lagrangian. Essentially we are assuming something like,
$\tvl S_0 \bvr \mspace{10mu}\sim \mspace{10mu}\bvl S\bvr,$ where $\tvr,$ and $S_0$ 
are respectively the true vacuum, and the conventional $S$ matrix. $S$ stands here
for the modified $S$-matrix, defined below.

\smallskip
\smallskip \textbf{3. The Vacuum Energy.}
\smallskip

 Because in the calculations for the amplitude, the interaction 
Lagrangian is used twice, we are led to multiply this one, by a factor proportional 
to $1/L,$ which results in the amplitude being proportional to $L^2.$ The dimensions of the 
Lagrangian must be left unchanged, however. This constraint fully determines the 
multiplicative factor $\cal{M},$ which can be no other than 
${\cal{M}} = (\hbar/mc)/L,$ a dimensionless quantity indeed.
Here $m$ is the mass of the electron, and $c,$ the velocity of light.\\
 In Eq.(4), the presence of the delta function $1/(2\pi)^4 \int d^4\xo e^{i(\xo(p/\hbar - p'/\hbar - k)}$ allows the integration over $k$ to be performed. Furthermore it is straightforward to show that ($Tr$ means the trace of the quantity in brackets),
\begin{equation}
Tr\{(\pbar + mc)\gamma^\alpha(\ppbar + mc)\gamma^\beta g_{\alpha\beta} \}
 = -16 (p^\mu p'_\mu - (mc)^2),
\end{equation}
clearly, a physically very appealing result. Finally the variables $p$ and $p'$ are 
transformed as follows $ p \rightarrow p mc,\mspace{5mu} p' \rightarrow p' mc,$ which
guaranties that the momentums $p$ and $p'$ are dimensionless.\\
 For the amplitude, calculated with the conventional interaction Lagrangian multiplied by $\cal{M},$ it is then found,
\begin{equation}
\bvl S\bvr = \frac{16i\alpha}{(2\pi)^7}\mspace{5mu} L^2 \mspace{5mu}\frac{(mc)^2}{(\hbar)^2}
\mspace{10mu}\cal{I}
\end{equation}
\begin{equation}
{\cal {I}}  = \int d^4 p\mspace{5mu} d^4 p'\mspace{5mu} \frac {p^\mu p'_\mu - 1}
{(p^2 + 1 + i\epsilon)(p'^2 - 1 + i\epsilon)((p-p')^2 +i\epsilon)}
\end{equation}
The amplitude must be dimensionless: in Eq.(6), $\alpha$ is associated with the
factor $e^2,$ in the interaction Lagrangian, and the presence of $ (mc/\hbar)^2$ is a consequence of the interaction Lagrangian being proportional to $L^2.$ \\      
We assume that the integral, $\cal I,$ is of order unity. Then, 
\begin{equation}
\mid \bvl S\bvr \mid^2 = \frac{(16\alpha)^2}{(2\pi)^{14}}\mspace{5mu} L^4 \mspace{5mu}
\frac{(mc)^4}{(\hbar)^{4}} 
\end{equation}
specifies the total number of events in the volume of integration, $L^4.$
 Therefore, the number of events per unit volume and unit time is given by,
\begin{equation}
{\cal{N}}= \frac{(16\alpha)^2}{(2\pi)^{14}}\mspace{5mu}c \mspace{5mu}
\frac{(mc)^4}{(\hbar)^{4}} =  1.23\mspace{5mu} 10^{39}{\rm cm}^{-3}{\rm s}^{-1}
\end{equation}

The contribution of particle-pair creation to the vacuum density is equal to
the pairs mass multiplied by {$\cal{N}$} and by the lifetime of the pairs,
\begin{equation}
\rho_V = 2m {\cal{N}} t_P = 1.2 \mspace{5mu} 10^{-31}{\rm g}\mspace{3mu}{\rm cm}^{-3}
\end{equation}
The observations (Riess et al. 1998, Peerlmutter et al. 1999)
show that, $\Omega_\wedge =\rho_v/\rho_{crit}\sim 0.7,$ with $\rho_{crit}=
3H_0^2/{8\pi} = 1.88\mspace{3mu}10^{-29}h^2 {\rm g}{\mspace{5mu}}{\rm cm}^{-3}$ (c.f. Hartle, 2003, Eq.18.32), here $H_0$ is the Hubble constant, and $h \sim 0.72.$ Then,
\begin{equation}
\rho'_V = 6.8\mspace{2mu}10^{-30}\mspace{2mu}{\rm g}\mspace{3mu}{\rm cm }^{-3}.
\end{equation}
The calculated value for the vacuum energy is in amazingly good agreement with the
observations.

To summarize: The Interaction Lagrangian is modified to enable the amplitude squared, to be 
interpreted as the number of events occurring in the volume of integration. Dimensional 
considerations reveal that the number of events per unit volume and unit time must be 
proportional to $\alpha^2 c \mspace{2mu}(mc/\hbar)^4.$ The proportionality-component
is estimated from Feynman diagrams calculated with the modified Lagrangian. It is very remarkable indeed that such apparently unrelated factors as: the electron mass; the contributions from Feynman diagrams; the Planck time!, could contribute to assign to 
$\rho_V$ a value in such a good approximation with the observations.

\smallskip
\smallskip \textbf{4. Appendix}
\smallskip

$ \hbar = 1.05\mspace{3mu} 10^{-27} {\rm g}\mspace{3mu}{\rm cm}^2 \mspace{3mu}{\rm s}^{-1},
\mspace{15mu}m = 9.11\mspace{3mu}10^{-28}\mspace{3mu}{\rm g},\mspace{15mu}
mc/\hbar = 2.59 \mspace{3mu} 10^{10}\mspace{3mu}{\rm cm}^{-1}$\\
$ \alpha = e^2 /4\pi \hbar c = 1/137.04, \mspace{15mu} 
t_P = 5.39\mspace{3mu} 10^{-44}\mspace{3mu}{\rm s}$

\smallskip
\smallskip \textbf{5. References.}
\smallskip

Dolgov, A.D., Sazhin, M.V., Zeldovich, Ya.B. (1990): {\it Modern Cosmology}, Editions 
Frontieres, Gif-sur-Yvette, France 
\smallskip

\smallskip
Hartle, J.B. (2003): {\it Gravity}, Addison Wesley, San Francisco, CA
\smallskip

\smallskip
Mandl, F., Shaw, G. (2010): {\it Quantum Field Theory}, John Wiley and Sons Ltd, U.K.
\smallskip

\smallskip
Perlmutter, S. et al. (1999): {\it Measurements of Omega and Lambda from 42 
High-Redshift Supernovae}, {\sl Ap.J.} {\bf 517}, 565-586
\smallskip

\smallskip
Riess, A.G. et al. (1998): {\it Observational Evidence from Supernovae for an Accelerating
Universe and Cosmological Constant}, {\sl Astron.J.} {\bf 116}, 1009-1038
\smallskip

\smallskip
Zee, A. (2010): {\it Quantum Field Theory in a Nutshell}, Princeton University Press, 
Princeton, New Jersey
\smallskip

\end{document}